\documentstyle[sprocl,epsf]{article}

\def\Journal#1#2#3#4{{#1} {\bf #2}, #3 (#4)}

\def\PRL{\em Phys. Rev. Lett.}
\def\PRD{{\em Phys. Rev.} D}

\def\be{\begin{equation}}
\def\ee{\end{equation}}
\def\bea{\begin{eqnarray}}
\def\eea{\end{eqnarray}}
\begin{document}
\vspace{-0.2cm}
\title{ ISOLATED PROMPT PHOTON CROSS SECTIONS\footnote
{Invited paper presented by E. L. Berger at DPF'96, 1996 Meeting of the 
Division of Particles and Fields of the American Physical Society, 
Minneapolis, MN, August 10-15, 1996.  Argonne report ANL-HEP-CP-96-77.  
This work was supported in part by the 
U. S. Department of Energy, Division of High Energy Physics, Contract 
No. W-31-109-ENG-38.}}
\author{ EDMOND L. BERGER$^1$, 
         XIAOFENG GUO$^2$ and JIANWEI QIU$^2$
 }

\address{$^1$High Energy Physics Division, Argonne National Laboratory,\\
Argonne, IL 60439, USA \\
         $^2$Department of Physics and Astronomy, Iowa State University,\\
Ames, IA 50011, USA }


\maketitle
\abstracts{We show that the conventionally defined partonic cross section 
for the production of isolated prompt photons is not an infrared safe quantity.
We work out the case of $e^+e^-\rightarrow\gamma + X$, and discuss 
implications for hadron reactions.}

%============== Text ==================================================

In $e^+e^-$ and in hadron-hadron reactions at collider energies,
{\it prompt} photons are observed and their cross sections are measured only
if the photons are relatively isolated in phase space.  Isolation is required  
to reduce various hadronic backgrounds including those from the 
electromagnetic decay of mesons, e.g., $\pi^o \rightarrow 2\gamma$.
The essence of isolation is that a cone of half-angle $\delta$ is drawn about 
the direction of the photon's momentum, and the isolated cross
section is defined for photons accompanied by less than a specified
amount of hadronic energy in the cone, e.g., $E_h^{\rm cone}\leq
E_{\rm max} = \epsilon_h E_{\gamma}$; $E_{\gamma}$ denotes the energy of the 
photon.  Theoretical predictions will therefore depend 
upon the additional parameters $\delta$ and $\epsilon_h$.  Isolation removes 
backgrounds, but it also reduces the signal.  For example, it reduces the 
contribution from processes in which the photon emerges from the 
long-distance {\it fragmentation} of quarks and gluons, themselves produced 
in short-distance hard collisions.  

%Electron-positron reactions offer a relatively clean environment for the study 
%of prompt photon production in hadronic final states.  Since there are no 
%complications from initial state hadrons, $e^+e^-\rightarrow\gamma + X$ is a 
%good process in which to examine quantum chromodynamics (QCD) predictions in 
%the final state, and the data may be a good source of information on 
%quark-to-photon and gluon-to-photon fragmentation functions.  These 
%fragmentation functions are, in turn, needed for predictions of photon yields 
%in hadronic collisions.  
Hard photons in $e^+e^-$ processes arise as QED bremsstrahlung from the 
initial beams, radiation that is directed along angles near 
$\theta_{\gamma} = 0\ {\rm and}\ \pi$, and as final state 
radiation from {\it direct} and {\it fragmentation} processes.  The final 
state radiation populates all angles, with a distribution having both 
transverse, $1 + {\rm cos}^2 \theta_{\gamma}$, and longitudinal 
components~\cite{BGQ1}.  

Much of the predictive power of perturbative QCD derives from 
factorization theorems~\cite{CSS}.  {\it Conventional} factorization expresses 
a physical quantity as the convolution of a partonic term  
with a nonperturbative long-distance matrix element, and it requires that the 
partonic term, calculated perturbatively order-by-order in the the strong 
coupling strength $\alpha_s$, have no infrared singularities.  Using 
$e^+e^-\rightarrow\gamma X$ as an example, we demonstrate that the 
perturbatively calculated partonic part for the isolated photon cross 
section is not infrared safe.  The infrared sensitivity shows up first in the 
next-to-leading order quark-to-photon fragmentation contribution~\cite{BGQ2}.  

For the quark fragmentation contributions, there are two sources of hadronic 
energy in the isolation cone:
a) from fragmentation of the quark itself, $E_{frag}$, and b) from 
non-fragmenting final-state partons, $E_{partons}^{\rm cone}$, that enter  
the cone.  When the maximum hadronic energy allowed in the
isolation cone is saturated by the fragmentation energy, $E_{\rm
max}=E_{frag}$, there is no allowance for energy in the cone from other
final-state partons.  In particular, if there is a gluon in the final
state, the phase space accessible to this gluon is restricted.  By
contrast, isolation does not affect the virtual gluon exchange loop 
contribution.  Therefore, for isolated photons, the infrared singularity 
from the virtual contribution is not canceled completely by the 
{\it restricted} gluon emission contribution~\cite{BGQ2}.

\begin{figure}
\vskip -2.9cm
\hbox{\epsfxsize6.8cm\epsffile{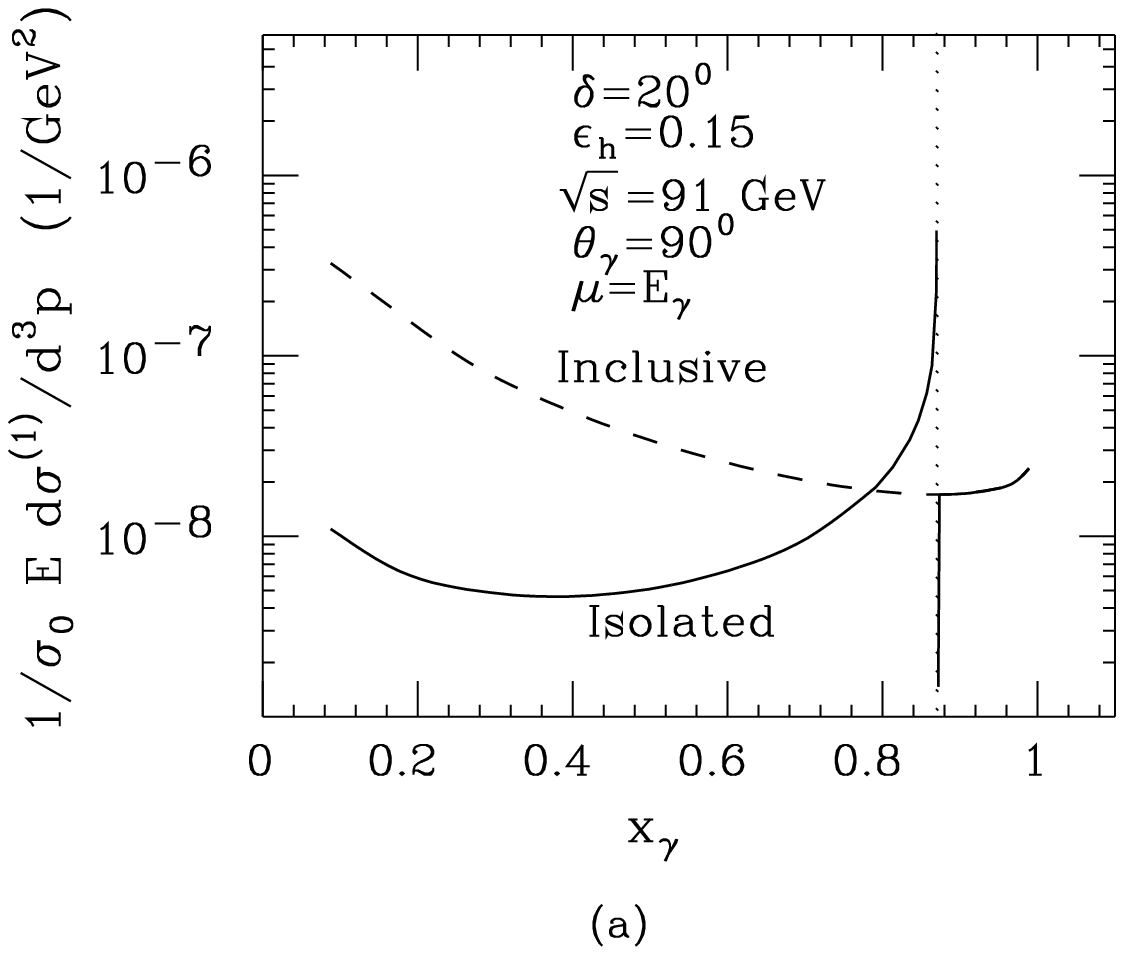}{\hskip -1.0cm}
\epsfxsize6.8cm\epsffile{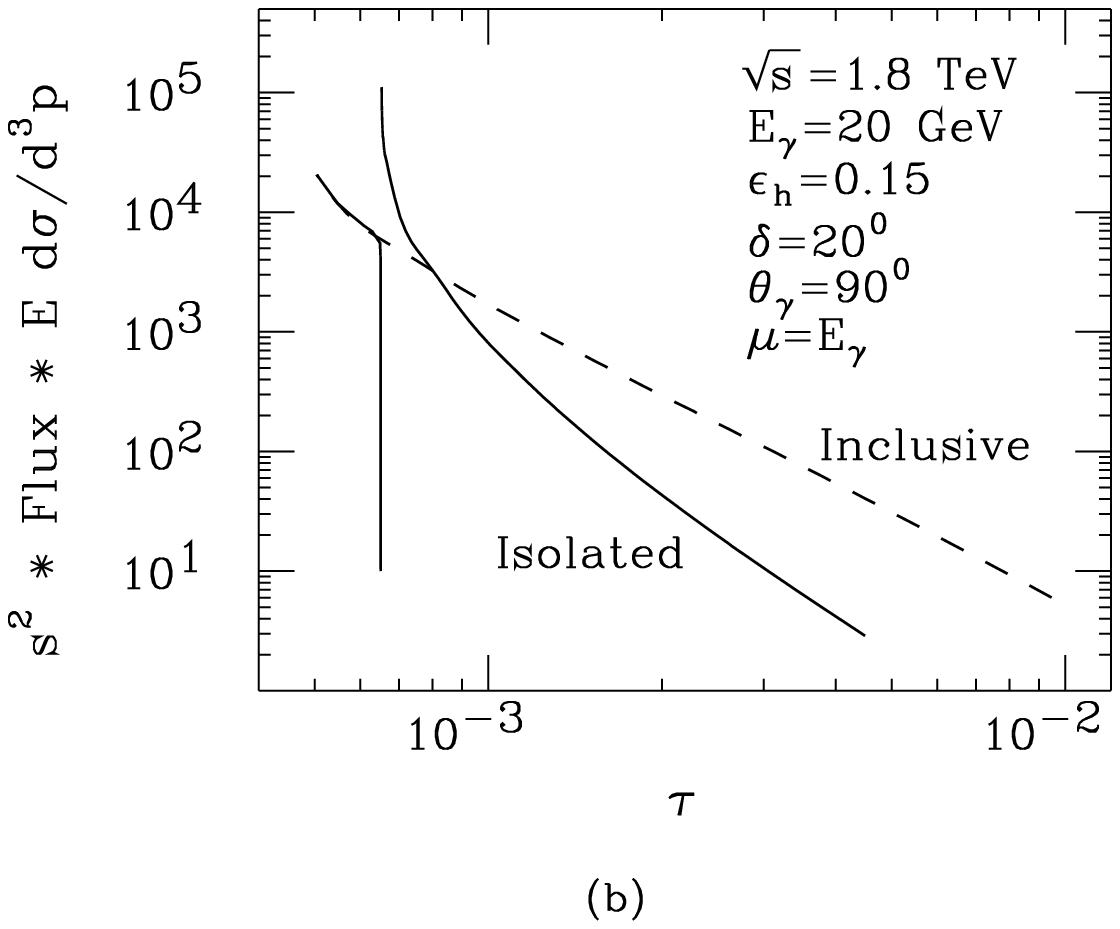}}
\caption{ One-loop quark fragmentation contributions to the isolated 
and inclusive cross sections (a) as a function of 
$x_\gamma = 2 E_\gamma / \protect\sqrt{s}$ in 
$e^+e^- \rightarrow \gamma X$ at $\protect\sqrt{s} = 91$~GeV, and (b) 
as a function of $\tau$ in $p \bar{p} \rightarrow \gamma X$
at $\protect\sqrt{s} =1.8$~TeV.}
\end{figure}
If conventional factorization were true, the fragmentation contributions to 
the physical cross section would be expressed in the 
factorized form
\begin{equation}
E_\gamma \frac{d\sigma^{iso}_{e^+e^- \rightarrow \gamma X}}{d^3\ell}
= \sum_c\  
    \int_{{\rm max}\left[x_\gamma,\frac{1}{1+\epsilon_h}\right]}^1 
         \frac{dz}{z}\
    E_c \frac{d\hat{\sigma}^{iso}_{e^+e^- \rightarrow cX}}{d^3 p_c} 
    \left(x_c=\frac{x_\gamma}{z}\right) 
    \frac{D_{c \rightarrow \gamma}(z,\delta)}{z}\ ;
\label{e1}
\end{equation}
$x_\gamma=2E_\gamma/\sqrt{s}$ and $x_c=2E_c/\sqrt{s}$.  The sum  
extends over $c = q,\bar{q}$ and $g$, and 
$D_{c\rightarrow\gamma}(z,\delta)$ is the
nonperturbative function that describes fragmentation of parton
``$c$'' into a photon.  The lower limit of integration results
from the isolation requirement with the assumption that all
fragmentation energy is in the isolation cone~\cite{BGQ2}.
Because of the isolation condition, the phase space constraints are
different in three regions: a) $x_\gamma < 1/(1+\epsilon_h)$, b)
$x_\gamma = 1/(1+\epsilon_h)$, and c) $x_\gamma > 1/(1+\epsilon_h)$. 
We summarize here the situation in the different regions and show that 
the next-to-leading order partonic term for quark fragmentation,
$E_q d\hat{\sigma}^{iso}_{e^+e^- \rightarrow qX}/d^3p_q$,
is infrared sensitive~\cite{BGQ2} {\it at and below} the point 
\hbox{$x_\gamma = 1/(1+\epsilon_h)$}. 

When $x_\gamma < 1/(1+\epsilon_h)$, subprocesses with two-body final
states do not contribute.  Therefore, there
is no leading-order quark (or antiquark) fragmentation contribution, 
and one-loop virtual diagrams do not contribute.
The well-known infrared pole singularity of the real gluon emission diagrams, 
having the form $1/(1-x_q)$ as $x_q=x_\gamma/z \rightarrow 1$, {\it remains 
uncanceled} in $\hat{\sigma}^{iso}_{e^+e^-\rightarrow qX}$.  After 
convolution with $D_{q\rightarrow \gamma}(z)$, this inverse power infrared 
sensitivity yields a logarithmic divergence proportional to 
$\ell n(1/x_\gamma -(1+\epsilon_h))$ in the cross section 
$\sigma^{iso}_{e^+e^- \rightarrow \gamma X}$.  As shown in Fig.~1(a), this 
means that the isolated cross section would become larger than the inclusive 
cross section in the vicinity of $x_\gamma \rightarrow 1/(1+\epsilon_h)$, a
result that is certainly not physical.  This infrared sensitivity in 
$\hat{\sigma}^{iso}_{e^+e^-\rightarrow qX}$ signals a clear breakdown of
conventional perturbative factorization.

When $x_\gamma = 1/(1+\epsilon_h)$, it is possible that 
$x_q=x_\gamma/z=1$.  Therefore, the one-loop virtual gluon
exchange diagrams, whose contributions are proportional to $\delta(1-x_q)$, do 
contribute.  However, isolation constraints limit the phase space accessible 
to gluon emission in the real subprocess, $e^+e^-\rightarrow q\bar{q}g$.  
Consequently, the infrared divergences in the real and virtual
contributions do not cancel completely in the isolated case, unlike
the inclusive case.  Working in $n=4-2\epsilon$ dimensions, we find~\cite{BGQ2}
\begin{equation}
E_q \frac{d\sigma^{iso}_{e^+e^-\rightarrow q X}}{d^3p_q} 
\sim
    \left\{\frac{1}{\epsilon^2}+\frac{1}{\epsilon} 
          \left(\frac{3}{2}-\ell n\frac{\delta^2}{4}\right)
         \right\}\, \delta(1-x_q)\ 
  + \mbox{finite terms} \ .
\label{e2}
\end{equation}
The presence of the uncanceled $1/\epsilon$ and $1/\epsilon^2$ terms means 
that the regulator $\epsilon$ cannot be set to 0.  Therefore, 
at $x_q=1$, corresponding to $x_\gamma = 1/(1+\epsilon_h)$, the partonic
term for quark fragmentation is infrared
divergent, and the perturbative calculation is not well-defined.
Conventional perturbative factorization again breaks down.

To recapitulate, in $e^+e^-\rightarrow\gamma + X$, the next-to-leading order 
partonic term associated with the quark fragmentation contribution 
is infrared sensitive when $x_\gamma \leq 1/(1+\epsilon_h)$.  
Conventional perturbative factorization of the cross section for isolated 
photon production in $e^+e^-$ annihilation breaks down in the 
neighborhood of $x_\gamma = 1/(1+\epsilon_h)$.  The isolated cross section, as 
usually defined, is not an infrared 
safe observable and cannot be calculated reliably in conventional perturbative 
QCD at and below $x_\gamma=1/(1+\epsilon_h)$.  

In hadron collisions, $A + B \rightarrow \gamma X$, we are interested in the 
production of isolated prompt photons as a function of the photon's 
transverse momentum, $p_T$.  At next-to-leading order in QCD, one must
include fragmentation at next-to-leading order.  At this level, 
problems analogous to those in $e^+e^-$
annihilation are also encountered in the hadronic case.  
As an example~\cite{BGQ2},
we consider the contribution from a quark-antiquark subprocess
in which the flavors of the initial and final quarks differ:  $q' +
\bar{q}' \rightarrow q + \bar{q} + g$, where $q$ fragments to
$\gamma$.  We take equal values for the incident parton momentum 
fractions, $x_a = x_b = x = \sqrt{\tau}$.  
In the translation to the hadronic case, the variable $x_\gamma$ 
becomes $\hat{x}_T$ where $\hat{x}_T = 2 p_T/\sqrt{\hat{s}} \sim
x_T/x$ with $x_T=2p_T/\sqrt{s}$.  

The special one-loop quark fragmentation contribution
to the observed cross section takes the form 
\begin{equation}
E_\gamma \frac{d\sigma_{AB \rightarrow \gamma X}}{d^3\ell} 
\sim  
\int_{x_T^2}^1 d\tau\ \Phi_{q'\bar{q}'}(\tau)\ 
E_\gamma \frac{d\sigma_{q'\bar{q}'\rightarrow \gamma
X}}{d^3\ell}(\tau) \quad + \quad \mbox{other subprocesses} \ . 
\label{e7st}
\end{equation}
In Fig.~1(b), we show the integrand in Eq.~(\ref{e7st})
obtained after convolution with the parton flux $\Phi(\tau)$. 
The corresponding contribution to the hadronic cross section
is the area under the curve in Fig.~1(b) from 
$x_T^2$ to 1.  It is evident that the convolution with
the parton flux substantially enhances the influence of the region of
infrared sensitivity.  The divergences above and below the point
$\hat{x}_T = 1/(1 + \epsilon_h)$ [or
$\sqrt{\tau}=x_T(1+\epsilon_h)$] are integrable logarithmic
divergences, and thus they yield a finite contribution if an
integral is done over all $\tau$.  However, we stress
that the perturbatively calculated one-loop partonic cross section
$E_q\, d\hat{\sigma}^{iso}_{q'\bar{q}'\rightarrow q X}$, 
has an inverse-power divergence as $x_q\rightarrow 1$ and has 
uncanceled $1/\epsilon$ poles in dimensional regularization~\cite{BGQ2}.  
The pole divergence for $\hat{x}_T < 1/(1 + \epsilon_h)$ becomes a logarithmic
divergence after convolution with a long-distance fragmentation function.  
Even though this logarithmic divergence near $\sqrt{\tau}=x_T(1+\epsilon_h)$ 
is integrable, it is certainly not correct to accept at face value a 
prediction in which a perturbatively calculated isolated cross section 
exceeds the inclusive cross section before the integration over $\tau$.  

The results in both the $e^+e^-$ and hadronic cases challenge us to find 
a modified factorization scheme and/or to devise more appropriate infrared
safe observables.

%%%%%%%%%%%%%% End of References %%%%%%%%%%%%%%%%%%%%%%%%%%%%%%%%%%%%%%%


\vspace{-0.4cm}
\section*{References}
\vspace{-0.2cm}
\begin{thebibliography}{99}

%%%%%%%%%%%%% Begin References %%%%%%%%%%%%%%%%%%%%%%%%%%%%%%%%%%%%%%%%

\bibitem{BGQ1}
 E. L. Berger, X. Guo, and J. Qiu, \Journal{\PRD} {53}
{1124} {1996}.

\bibitem{CSS}
 J. C. Collins, D. E. Soper and G. Sterman, in {\it Perturbative
 Quantum Chromodynamics}, ed. A. H. Mueller 
 (World Scientific, Singapore, 1989).

\bibitem{BGQ2} 
 E. L. Berger, X. Guo, and J. Qiu, \Journal{\PRL}{76}{2234}{1996}; 
hep-ph/9605324, {\it Phys. Rev.} D {\bf 54} (1996), in press.

\end{thebibliography}
\end{document}